\documentclass[11pt]{iopart}

\usepackage{graphicx}
\usepackage{subfigure}

\usepackage[bf]{caption} 

\usepackage{setspace}

\usepackage{cite}

\begin{document}

\subfiglabelskip=0pt

\title[]{Effect of particle contact on the electrical performance of NTC-epoxy composite thermistors}

\author{D. B. Deutz$^{1,2*}$, S. van der Zwaag$^{1}$ and P. Groen$^{1,3}$}

\address{$^1$ Novel Aerospace Materials Group, Faculty of Aerospace Engineering, Delft University of Technology, Kluyverweg 1, 2629HS, Delft, the Netherlands}
\address{$^2$ University Library, University of Southern Denmark, Campusvej 55, 5230, Odense, Denmark}
\address{$^3$ Holst Centre, TNO, High Tech Campus 31, 5605KN Eindhoven, the Netherlands}
\ead{dbd@bib.sdu.dk}
\vspace{10pt}
\begin{indented}
\item[]November 2019
\end{indented}

\begin{abstract}
As demand rises for flexible electronics, traditionally prepared sintered ceramic sensors must be transformed into fully new sensor materials that can bend and flex in use and integration. Negative temperature coefficient of resistance (NTC) ceramic thermistors are preferred temperature sensors for their high accuracy and excellent stability, yet their high stiffness and high temperature fabrication process limits their use in flexible electronics. Here, a low stiffness thermistor based on NTC ceramic particles of micron size embedded in an epoxy polymer matrix is reported. The effect of particle-to-particle contact on electrical performance is studied by arranging the NTC particles in the composite films in one of three ways: 1) Low particle contact, 2) Improved particle contact perpendicular to the electrodes and 3) dispersing high particle contact agglomerated clumps throughout the polymer. At 50 vol.\% of agglomerated NTC particles, the composite films exhibit a $\beta$-value of 2069 K and a resistivity, $\rho$, of 3.3$\cdot 10^5$  $\Omega$m, 4 orders of magnitude lower than a randomly dispersed composite at identical volume. A quantitative analysis shows that attaining a predominantly parallel connectivity of the NTC particles and polymer is a key parameter in determining the electrical performance of the composite film.
\end{abstract}


\vspace{2pc}

\vspace{2pc}
\submitto{\SMS} 

\clearpage

\section{Introduction}
\doublespacing

Demand for flexible sensors is rising with applications ranging from implantable medical equipment to robotic skins to consumer electronics \cite{Amjadi2016,Han2017,Servati2017}. Accurate temperature sensing is vital to the operation of these electronics. Traditionally the leading methods to measure temperature have been thermocouples, resistance temperature detectors (RTDs) and negative temperature coefficient of resistance (NTC) ceramic sensors. Thermocouples and RTD's typically have low sensitivity (at +0.1\% change in resistance per $^\circ$C), making NTC based sensors the preferred choice in industry for their high accuracy (-4\% change in resistance per $^\circ$C) and stability \cite{Strouse2008,Feteira2009,Kanao2015}. Molding these rigid sensors onto flexible substrates is a challenge \cite{Someya2005,Yao2015,Yokotaa2015,Kanao2015}, and a large number of alternative methods for measuring temperature have been suggested, yet poor stability has continued to limit their use\cite{Mclachlan1990,Strumpler1996,Yu2009,Bielska2009,Huang2013,Jeon2013,Khan2016,Billard2016,Li2017}. 

Mixing stable NTC particles into a polymer matrix to embue a printable polymer with high accuracy NTC sensing could be a prudent alternative. In 2019, Katerinopoulou et al. demonstrated a printed NTC composite thermistor with electrical properties near to a sintered NTC ceramic sensor \cite{Katerinopoulou2019}. This was achieved by mixing a high volume content, over 50 \%, of the inhomogeneous ceramic filler in a polymer binder. At such high volume content, it is likely the ceramic filler is above the percolation threshold and conductive paths form over the thickness of the composite \cite{Miyasaka1982,Foulger1998,Aneli2013}. The question then remains whether such high volume content is required, or merely the formation of conductive paths. To examine this more closely, here we prepare three types of NTC-polymer composite films with increasing degrees of particle-to-particle contact. We then vary the NTC volume content of each composite type to assess the influence of volume faction and particle contact on the electrical performance. 

This paper is organized as follows. First the methods to fabricate composite films of NTC ceramic particles dispersed in a polymer matrix for the three degrees of particle contact are described. Then the electrical performance of the three types of composite films is analyzed by measuring the resistivity, $\rho$, and the $\beta$-value. The achieved values are compared to an analytical model and the effective interparticle distance of each composite type is evaluated. We argue that the degree of parallel particle-to-particle contact of the NTC particles within the composite films dominates the electrical performance in these NTC ceramic-polymer sensors. 

\section{Materials and methods}

\noindent The NTC powder used in this work has the composition Mn$_{2.45}$Ni$_{0.55}$O$_4$. Sintered NTC ceramic thermistors of this composition are charactarized with a resistivity at 25 $^{o}$C, $\rho_{25}$, on the order of 5$\cdot 10^{3}$ $\Omega\cdot$m and a $\beta$-value of 3800 K. The NTC powder was prepared from mixtures of Mn$_2$O$_3$ and NiO in the appropriate ratios. The powders were ball-milled on a roll bench (Gladstone Engineering Co. Ltd) using 2 mm yttria-stabilized ZrO$_2$ balls in isopropanol and calcined in a Nabertherm high temperature furnace at 800 $^{o}$C for two hours, 950 $^{o}$C for 4 hours (heated at a rate of 350 K/h) and then allowed to cool to room temperature to develop the desired tetragonal spinel structure. After calcination, the agglomerated powder was milled with 5 mm yttria-stabilized ZrO$_2$ balls in cyclohexane on a roll bench for 16 hours. The particle size distribution was measured by laser diffraction (Beckman Coulter Laser Diffraction Particle Size Analyzer) to be between 1 $\mu$m and 10 $\mu$m, with remaining agglomerates of size distribution 40 $\mu$m to 250 $\mu$m. The largest agglomerates were removed by sieving through a 63 $\mu$m sieve. The structure of the tetragonal spinel NTC powder was checked after milling by X-ray diffraction (Bruker D8 Advance diffractometer) using Cu-K$\alpha$ (Figure \ref{fig:figS1}). \\

\noindent To measure the electrical properties of the NTC-epoxy composite films, Au electrodes were applied on either side of the films with a sputter coater (Balzers Union, SCD 040). The films were post cured and dried at 100 $^{o}$C for 1 hour before any measurements took place. To calculate the resistivity, $\rho$, DC resistance measurements were performed on disk shaped sensors (9 mm diameter and 1 mm thickness) at 25 $^\circ$C using a high resistance meter (Agilent 4339B) in combination with a component test fixture (Agilent 16339A) set to a voltage of 10 V. The resistivity, $\rho$, was then calculated from: $ \rho= R_{DC}\cdot A/t$, where $R_{DC}$ is the DC resistance, $A$ is the electrode area, and $t$ is the film thickness. To calculate the $\beta$-value, the resistance was measured with an AC resistance meter (HP 4276A, LCZ meter) set to a frequency of 1 kHz and a voltage of 1 V, at 25 $^{o}$C and 85 $^{o}$C in a water-cooled oil bath (Julabo, SE Class III, 12876). The $\beta$-value was then calculated from: $\beta = ln(R_{25}/R_{85}) / ((1/T_{25}) - (1/T_{85}))$, where $R_{25}$ and $R_{85}$ are the AC resistance at 25 and 85 $^o$C, respectively, and $T_{25}$ and $T_{85}$ are the temperature in K at 25 and 85 $^o$C, respectively. Impedance measurements were performed using a potentiostat/galvanostat (Autolab PGSTAT 302 N) coupled to a frequency analyzer with a 10 mV (rms) sinusoidal perturbation with respect to the open circuit potential. The microstructure of the composite films was observed using a Field Emission - Scanning Electron Microscope (SEM) (JEOL, JSM-7500F) operated in backscattered electron (composition) mode. The data underpinning this work is freely available at DOI:10.5281/zenodo.3243167.

\section{Fabrication of NTC-epoxy composite films}

\begin{figure}[ht!]
	\captionsetup{format=plain}
	\centering
	\includegraphics[width=0.70\textwidth]{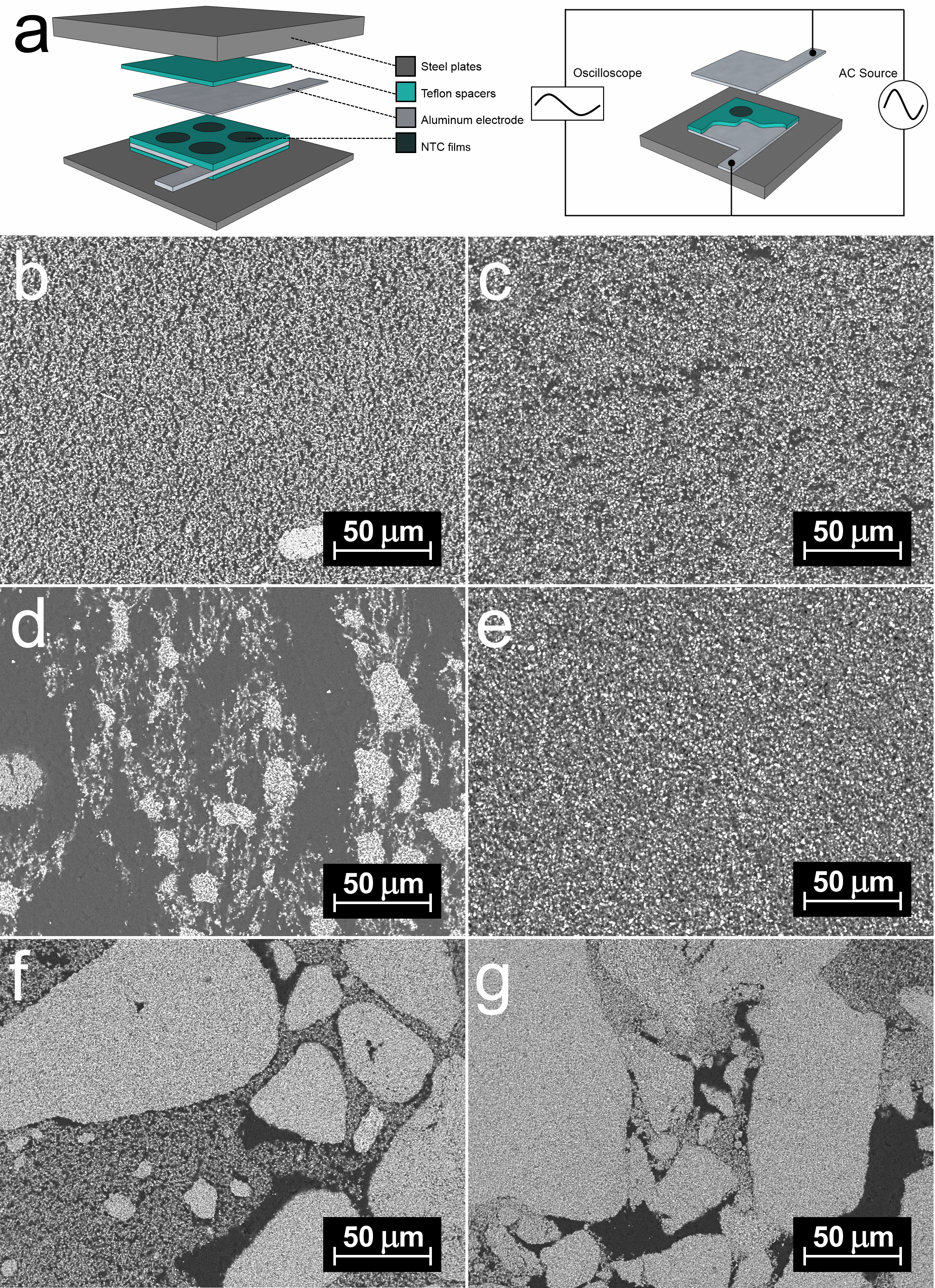}
	\caption{\label{fig:fig1} Fabrication and microstructure of NTC-epoxy composite films.   (a) Schematic of the mold used to fabricate all NTC-epoxy composite films (left) and the adjusted setup of the mold to induce dielectrophoretic alignment of NTC particles within the composite films (right). (b) Scanning electron micrograph of a cross section of a cured, randomly dispersed NTC-epoxy 15 vol.\% composite, and (c) a 30 vol.\% composite. (d) Scanning electron micrograph of a cross section of a cured and aligned NTC-epoxy 15 vol.\% composite, and (e) a 30 vol.\% composite. (f) Scanning electron micrograph of a cross section of a cured and agglomerated NTC-epoxy 35 vol.\% composite, and (g) a 50 vol.\% composite.}
\end{figure}

\noindent NTC-epoxy composite films were prepared by mixing NTC particles in an optically clear two-component epoxy polymer, (Epotek 302-3M, Epoxy Technology Inc., Billerica, Ma, USA), at a volume of 0 to 50, in a planetary speed mixer (DAC 150 FVZ, Hauschild, Germany). The slurry was degassed, poured into a prepared Teflon mold and clamped between two steel plates, as schematically depicted in Figure \ref{fig:fig1}a.
Three types of composite films were prepared as follows:  

\begin{enumerate}
	\item Low to no particle-to-particle contact: Random composite films -- prepared by randomly dispersing sieved NTC particles with a particle size between 1 $\mu$m and 10 $\mu$m in the uncured polymer and cured overnight at 50 $^{o}$C.  
	\item  Improved parallel particle-to-particle contact: Aligned composite films -- prepared by aligning sieved NTC particles in the uncured polymer by dielectrophoresis \cite{Deutz2017b} at room temperature for 3 hours (See Figure \ref{fig:figS2} for a top view of the particle alignment over time). The composite was cured overnight at 50 $^{o}$C with the alignment field turned on.
	\item Improved particle-to-particle contact overall: Agglomerated composite films -- prepared by mixing agglomerated NTC particles in the uncured polymer and cured overnight at 50 $^{o}$C. This agglomerated powder was reserved after the milling step and was not sieved.  
\end{enumerate}

\noindent The dispersion of the NTC particles in the three types of composite films can be inferred from the SEM micrographs of Figure \ref{fig:fig1}b - g. In the random composite films the micron sized NTC particles are homogeneously dispersed in the epoxy polymer at 15 vol.\% (Figure \ref{fig:fig1}b) and 30 vol.\% (Figure \ref{fig:fig1}c). In the 15 vol.\% aligned films, the micron sized NTC particles are aligned in parallel clumps between the electrodes (Figure \ref{fig:fig1}d), while at 30 vol.\% the dielectropherisis field has had no effect on the internal structure of the film (Figure \ref{fig:fig1}e). Aligning particles by dielectrophoresis is significantly more effective at low volume contents of filler, as the particles need space to move and link up into chains \cite{Deutz2017b, Bowen1998, Wilson2005, vandenEnde2010}. In the agglomerated films (Figure \ref{fig:fig1}f,g) large, $>100\mu$m agglomerated clumps of NTC particles are bounded by areas with no NTC particles and areas with well dispersed micron sized NTC particles. In the 50 vol.\% composite films connected agglomerations form a continuous path from one side of the film to the other. In the 35 vol.\% films the path is disrupted by areas with well dispersed, non-connected NTC particles. 

\section{Electrical performance of composite films}

\begin{figure}[ht!]
	\captionsetup{format=plain}
	\centering
	\includegraphics[width=0.80\textwidth]{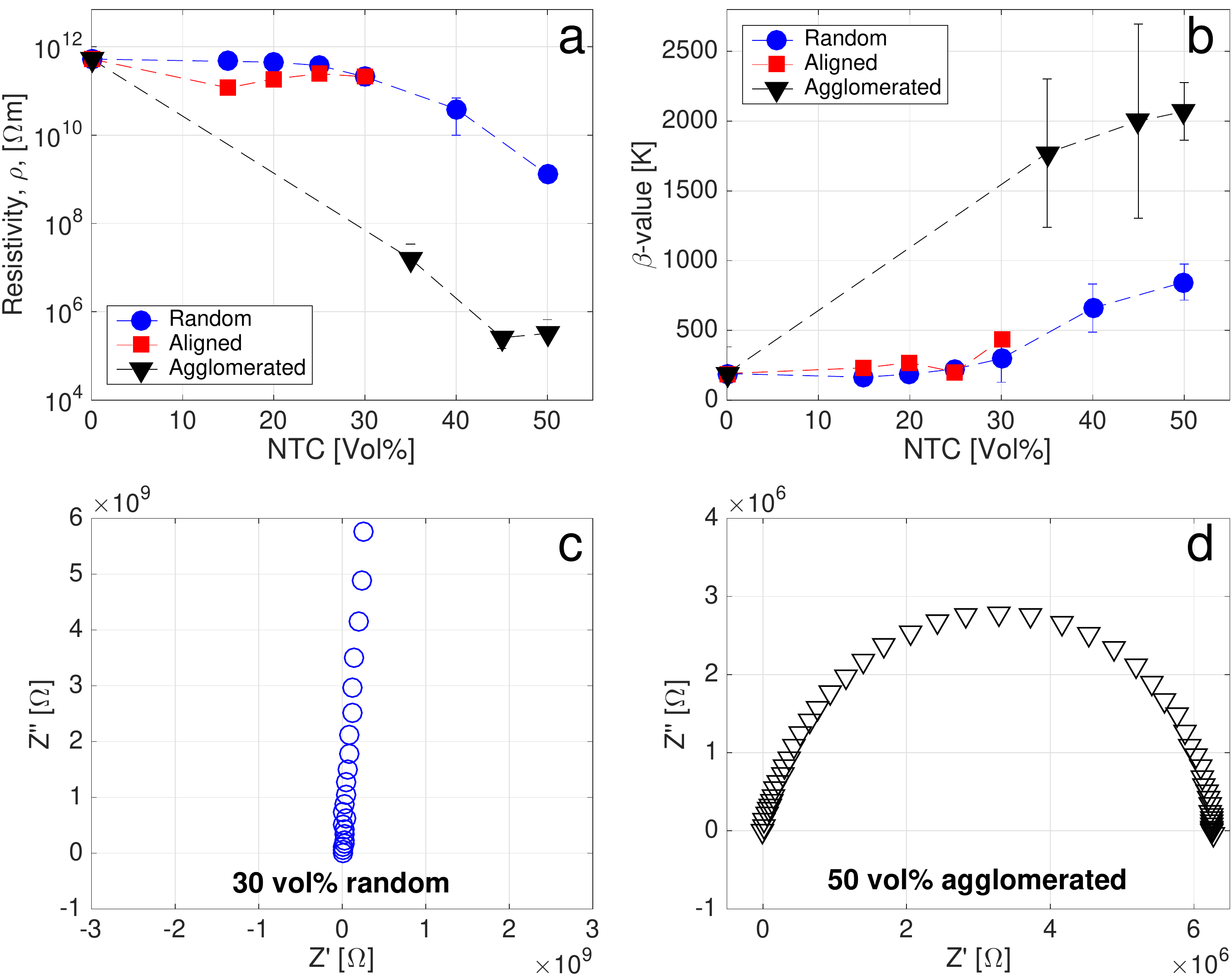}
	\caption{\label{fig:fig2} Electrical properties of NTC-epoxy composite films.   (a) Resistivity, $\rho$, as a function of NTC content.  (b) $\beta$-value as a function of NTC content. (c) Nyquist plot of a 30 vol\% randomly dispersed composite film. (d) Nyquist plot of a 50 vol\% agglomerated composite film. Blue circles represent randomly dispersed composite films. Red squares represent dielectrophoretically aligned composite films. Black triangles represent agglomerated composite films. Dashed lines are a guide to the eye. }
\end{figure}

Figure \ref{fig:fig2}a represents the calculated resisitivty, $\rho$, at 25$^\circ$C as a function of NTC content. Blue circles represent the values extracted for films with randomly dispersed particles. The films are hardly conductive, dominated by the resistivity of the matrix. Composite films with aligned NTC particles, represented by the red squares, show some marginal improvement in resistivity, as compared to random composites. The difference decreases with increasing NTC content. However, the resistivity dramatically decreases when agglomerated clumps of NTC particles are introduced, as shown by the black triangles. This switch from capacitive to conductive performance is supported by the Nyquist plots shown in Figure \ref{fig:fig2}c,d. The $\beta$-value as a function of NTC content is presented in Figure \ref{fig:fig2}b. Here too, the performance of the aligned and random composites is dominated by the polymer matrix. The agglomerated composites attain a $\beta$ of 2069 K at an NTC content of 50 vol.\%. This value may only be half of that obtained in sintered nickel manganite NTC ceramics, yet the composites can be simply integrated in flexible electronics \cite{Katerinopoulou2019}.

\clearpage

\section{Particle contact and interparticle distance in the composite films}

The explanation for the decrease in resistivity, $\rho$, and increase in $\beta$-value of the agglomerated NTC-epoxy composite films strongly depends on the particle contact. There are many models to describe the dependence of the electrical performance of ceramic-polymer composites on the content and connectivity of the phases \cite{Mclachlan1990, Bruggeman1935, Ruschau1992, Youngs2002, Almond2006}. Here we use an adjusted version of the rule of logarthimic mixing, or Lichtenecker model, to describe the dependence of the resistivity, $\rho$, on volume content and connectivity, given by Eq. 1 as \cite{Lichtenecker1926}:  

\begin{equation}
\rho^{n}_{c} = \varphi\rho^n_{NTC} + (1-\varphi)\rho^n_{p}
\label{eq:Licht}
\end{equation}

\begin{figure}[hb!]
	\captionsetup{format=plain}
	\centering
	\includegraphics[width=0.80\textwidth]{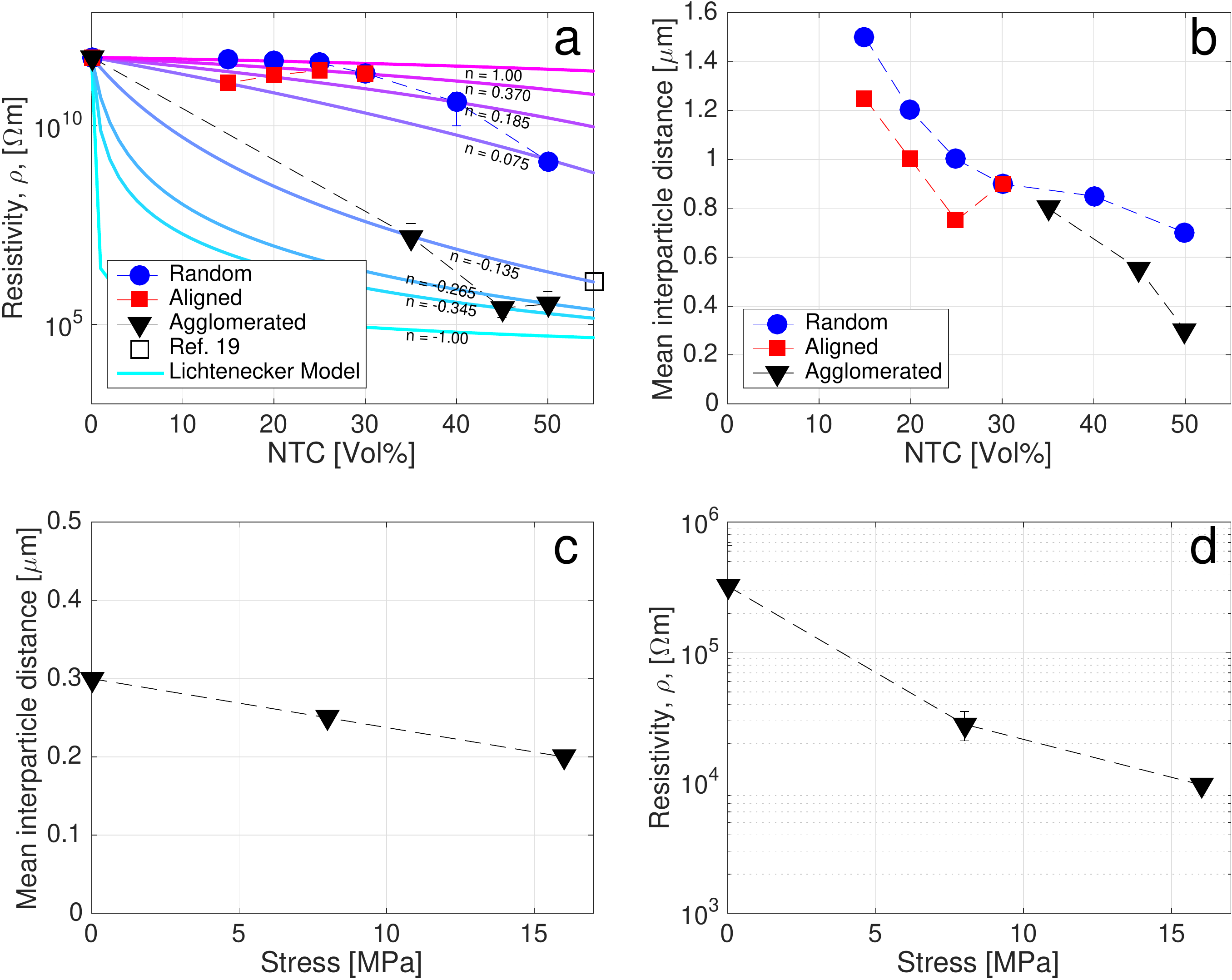}
	\caption{\label{fig:fig3}Effect of particle connectivity on the resistivity of NTC-epoxy composite films. (a) Lichtenecker model superimposed over the plot of resistivity, $\rho$, as a function of NTC content. (b) Mean interparticle distance as a function of NTC content. (c) Resistivity, $\rho$, as a function of applied uniaxial stress, $\sigma$. (d) Mean interparticle distance as a function of applied uniaxial stress, $\sigma$. Blue circles represent randomly dispersed composite films. Red squares represent dielectrophoretically aligned composite films. Black triangles represent agglomerated composite films. Full lines represent the Lichtenecker connectivity model, varying from the lower bound parallel connectivity (shown in cyan) to the upper bound series connectivity (shown in magenta). Dashed lines are a guide to the eye. }
\end{figure}

\noindent Where $\varphi$ is the volume content of NTC particles in the composite film, $\rho_p$ is the resistivity of the insulating polymer matrix, $\rho_{NTC}$ is the resistivity of the NTC particles, and $\rho_{c}$ is the resistivity of the composite. The Lichtenecker model bridges the gap between a perfect series and perfect parallel structure within a composite by varying the factor $n$ from -1 (for parallel addition) to +1 (for series addition). Figure \ref{fig:fig3}a shows the Lichtenecker model superimposed over the resistivity as a function of NTC content (shown in Figure \ref{fig:fig2}a). The upper bound (shown in magenta) represents a perfect series connectivity of the two phases of the composite film, while the lower bound (shown in cyan) represents perfect parallel connectivity, or perfect contact from one particle to the next. \\

\noindent In random composite films, even at high NTC content, the particles are mainly in series with the polymer. Quantitative fits of the $n$ values range from +1, perfect series connectivity, to +0.075. Aligned films do show some improvement towards parallel connectivity, yet the internal structure is still predominantly series. Only in the agglomerated films is some degree of particle-to-particle contact achieved, with quantitative fits of $n$ ranging from -0.135 to -0.345. For comparison, we have included the printed NTC sensor reported by Katerinopoulou et al. \cite{Katerinopoulou2019}, with extracted approximate values for the resistivity of the sensor as 1.2 $\cdot 10^6\Omega$m at an NTC volume content of 55 \%. These values place the printed NTC sensor \cite{Katerinopoulou2019}, composed of randomly dispersed NTC particles, neatly on the parallel connectivity side of the Lichtenecker model with a quantitative fit for $n$ of -0.135.\\

\clearpage

\noindent To confirm that the resistivity is dominated by particle-to-particle contact we turn to the mean interparticle distance within the composite films, presented in Figure \ref{fig:fig3}b as a function of NTC content. Micrographs of each composite film were taken at 1500x magnification and thresholded until only the particles in the foreground of the image remained. Incomplete particles (on the edge of the image) and particles smaller than  1 $\mu$m  were then removed from the binary image, holes inside each particle were filled, and the boundary of each remaining particle was smoothed. The centroid of each remaining particle, the particle size, and the pair-wise distance from each centroid to the next were calculated. The mean interparticle distance was taken from a histogram of all the nearest neighbor distances of each particle centroid. \\

\noindent The mean interparticle distance decreases as a function of NTC content. At low NTC content aligning NTC particles decreases the interparticle distance with respect to random composite films. At 30 vol.\% NTC content, dielectrophoretic processing had no measurable effect on the aligned composites as both the interparticle distance and the electrical performance is identical to the values measured for random composites. At high NTC content the agglomerated films have a lower interparticle distance than all random and aligned films. Even though there is a clear spacing between particle clusters (Figure \ref{fig:fig1}c), the particles within each cluster are packed so closely that the mean interparticle distance remains low. Interestingly, composite films around the 0.8 $\mu$m range have a wide spread in resistivity. At 35 vol.\% the agglomerated films have an interparticle distance that is higher than both 25 vol.\% aligned films and 50 vol.\% random films, indicating that, on average, even though the particles are farther apart the resistivity of the agglomerated film is 2 - 4 orders of magnitude lower. \\

\noindent Aligning the particles only marginally improved the resistance, even though aligned composites do indeed have a lower mean interparticle distance than random composites (Figure \ref{fig:fig3}a,b). In piezoelectric sensors dielectrophoretic alignment significantly improves the performance of low volume content composite sensors. The key piezoelectric sensing parameter $d_{ij}$, the piezoelectric charge coefficient, is proportional only to the charge at the electrodes, $Q$, over the change in applied force, $\Delta F$ \cite{Bowen1998, vandenEnde2010, Deutz2017b}. When the current, $I$, is the driving factor in sensing, as it is for NTC and pyroelectric composites (where the pyroelectric coefficient, $p$, scales with $I$: $p = I\cdot A\cdot H$ and $H$ is the constant heating rate) dielectrophoretic particle alignment has only a muted effect on sensing \cite{Khanbareh2014}. We suggest that dielectrophoretic particle alignment will only be effective for composite sensors in cases where the key sensing parameter is independent of the current, $I$, and particle-to-particle contact is not required. \\

\noindent To decrease the mean interparticle distance independent of the particle contact, agglomerated 50 vol.\% NTC composite films were subjected to increasing amounts of uniaxial pressure in a hot press (Simplimet II, Buehler Ltd.) set to 130 $^o$C for 1 hour. The resistivity at 25 $^o$C and mean interparticle distance of the films is presented in Figure \ref{fig:fig3}c and d as a function of the applied stress, $\sigma$. The resistivity decreases monotonically with the mean interparticle distance by approximately an order of magnitude per 0.1 $\mu$m. Yet, in random composites from a volume content of 15 to 25, the interparticle distance decreases by approximately 0.5 $\mu$m (Figure \ref{fig:fig3}b) without any significant reduction in resistivity. This indicates that, while interparticle distance does have an effect on resistivity, it is the particle contact and NTC content that are the key parameters in determining electrical performance of the NTC composite thermistor.

\section{Conclusions}
Random, aligned and agglomerated NTC ceramic - epoxy polymer composite films were fabricated to investigate the effect of particle contact on the resistivity, $\rho$, and $\beta$-value of printed NTC composite thermistors. Regardless of the volume of NTC particles in the composite films, the resistivity of both randomly dispersed and aligned particle composites are dominated by the resistivity of the polymer matrix. Only in composite films with large agglomerated NTC particles, over $10$\% of the film thickness in size, does the electrical performance begin to approach that of sintered NTC ceramics. Quantitative analysis has shown that the origin is an increase in parallel connectivity of the NTC particles and epoxy polymer matrix. While the interparticle distance decreases monotonically with NTC content and resistivity, the effect on the conductivity is minor when compared to that of the particle contact.

\ack
We would like to thank Sophie Schuurman of Vishay BC components and Sophie Fritsch of CIRIMAT CNRS for stimulating discussions. 

\section*{References}
\bibliography{references}

\clearpage
\section{Supplemental information}
\setcounter{figure}{0}
\renewcommand\thefigure{S\arabic{figure}} 
 
\begin{figure}[ht!]
	\captionsetup{format=plain}
	\centering
	\includegraphics[width=0.90\textwidth]{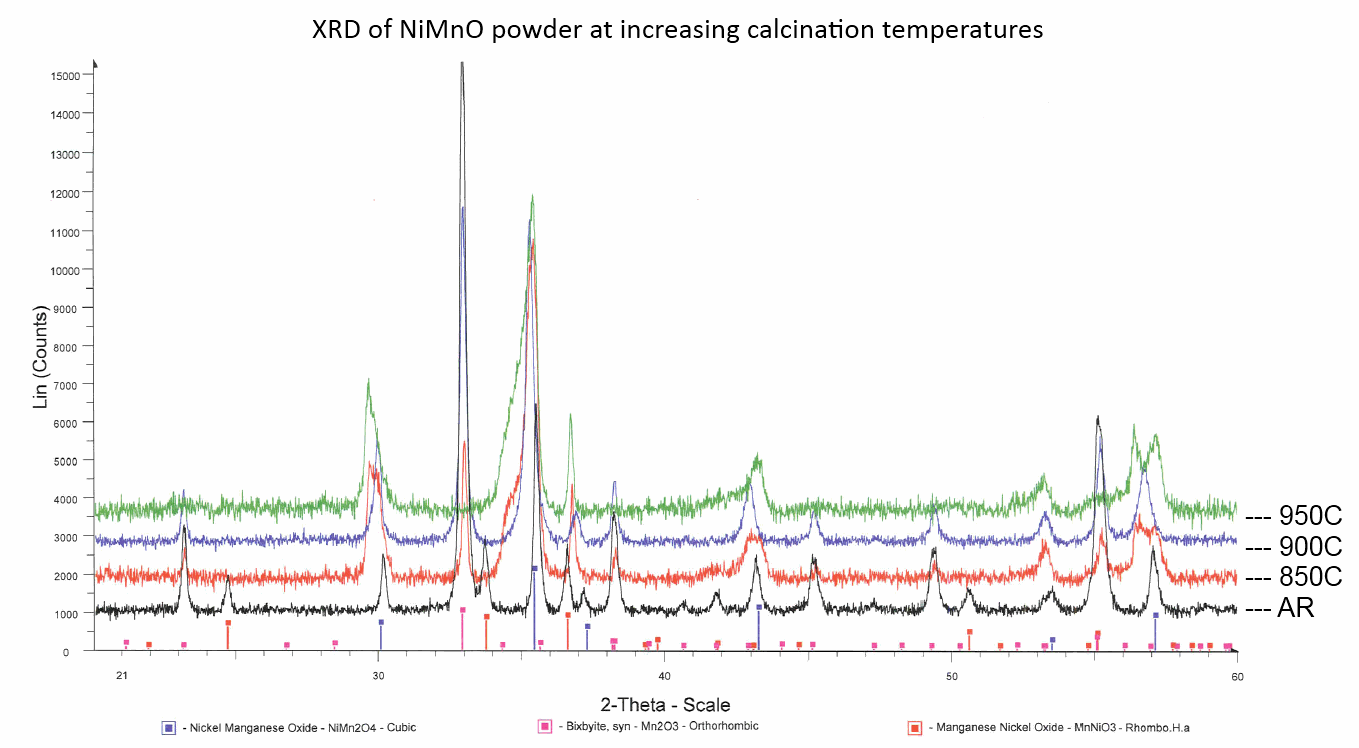}
	\caption{\label{fig:figS1} X-ray diffraction pattern of the as-received (AR) Nickel manganite powder, and the structure after processing at 850, 900 and 950 $^o$C. }
\end{figure}

 \begin{figure}[ht!]
	\captionsetup{format=plain}
	\centering
	\includegraphics[width=0.90\textwidth]{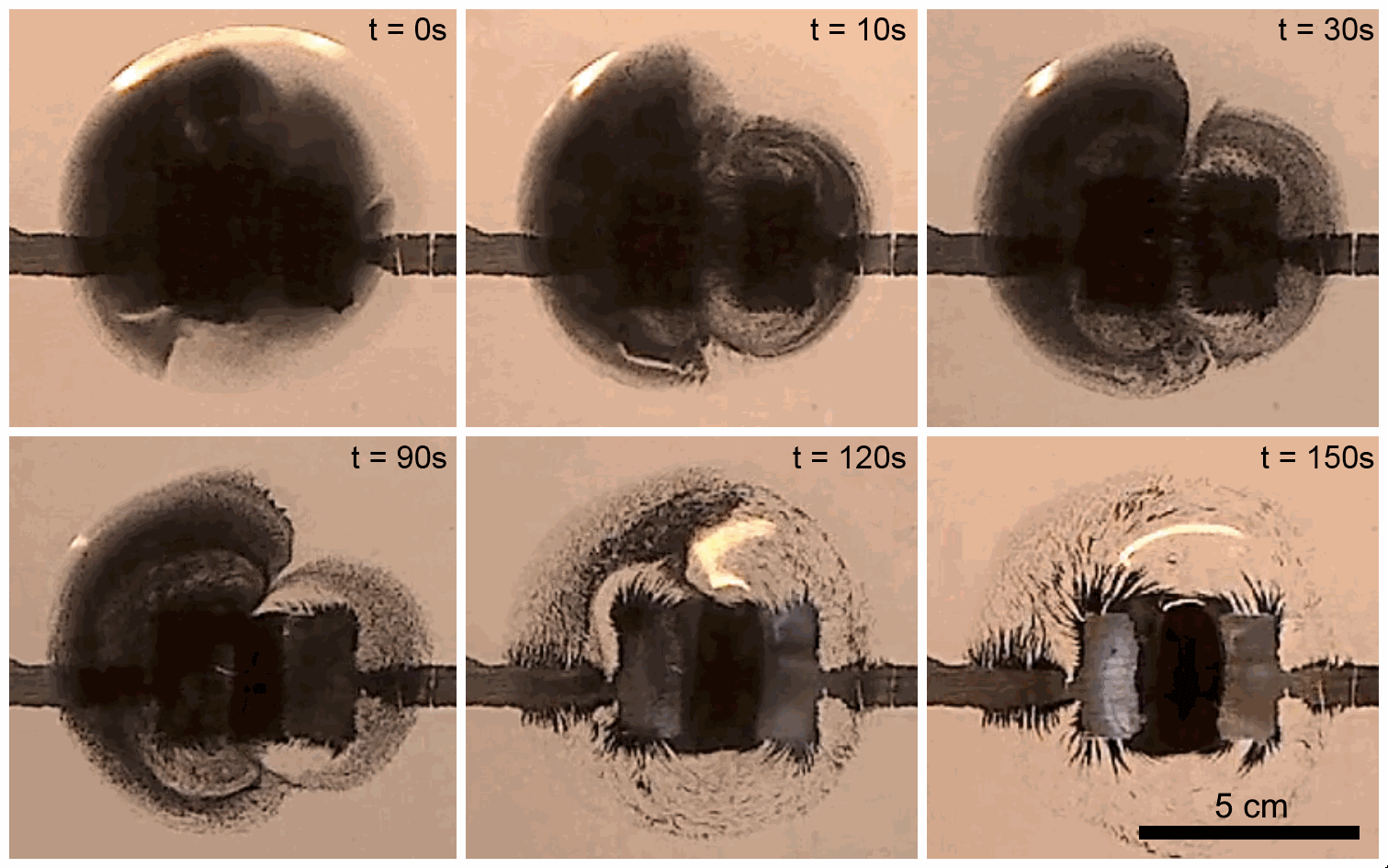}
	\caption{\label{fig:figS2} Top view of the alignment of NTC particles in the uncured epoxy polymer by dielectrophoresis, over time. The distance between the two silver, rectangular shaped electrodes is 2 mm. }
\end{figure}

\end{document}